\documentclass[twocolumn,aps]{revtex4}
\include{epsfig}

\newcommand{\be}{\begin{equation}}
\newcommand{\ee}{\end{equation}}
\newcommand{\ba}{\begin{eqnarray}}
\newcommand{\ea}{\end{eqnarray}}
\newcommand{\GeV}{\mbox{GeV}}
\newcommand{\MeV}{\mbox{MeV}}

\usepackage{graphicx}
\usepackage{bm}

\begin{document}

\title{
 Instanton-induced Azimuthal Spin Asymmetry in Deep Inelastic Scattering}
\author{Dmitry Ostrovsky} 
\author{Edward Shuryak}
\affiliation{Department of Physics and Astronomy,\\ State
University of New York at Stony Brook, New York 11794, USA }
\date{\today}

\begin{abstract}
\noindent 
It is by now well understood that spin asymmetry in deep inelastic
scattering (DIS) can appear if two things are both
present: (i) a chirality flip of the struck quark; (ii)  
a nonzero T-odd phase due to 
its final state interaction. So far (i) was attributed to a new
structure/wave function of the nucleon and (ii) to some gluon exchanges.
We propose a new mechanism  utilizing strong vacuum fluctuations
of the gluon field described  
semiclasically by instantons, and show that both (i) and (ii) are present.
The magnitude of the effect is estimated using known parameters of the
 instanton ensemble in the QCD vacuum and known structure and 
fragmentation functions, without any new free parameters. The result
 agrees in sign and (roughly) in magnitude
with the available data on
 single particle inclusive DIS. Furthermore, our
predictions uniquely relate effects for
 longitudinally and transversely polarized targets.
\end{abstract}
\maketitle


\section{Introduction}

Perturbative QCD is well known to account correctly for the 
dependence
of structure and fragmention functions on the hard scale $Q^2$. 
On the other hand, 
 the conventional perturbative cascade of gluon emission and of the quark
pairs production (on which it is based) is clearly inadequate to
explain the original quark sea and glue itself, at a border scale
to nonperturbative regime $\mu\sim 1 \, GeV$. A lot of puzzles
about the nucleon structure are
revealed by experiments, and we understand neither their dynamical
origin nor  their magnitude  for other hadrons.

 Already at the inclusive
level of leptonic deep inelastic
scattering (DIS), we learned from the spin-$independent$ 
 structure functions that
the quark sea is rather strongly flavor polarized. 
The spin-$dependent$ DIS have further shown that the sea quarks are also
strongly
polarized in spin, 
in the direction  $opposite$ to the polarization of the nucleon (and the
valence
quarks): thus the ``spin crisis''. 

Significant experimental efforts
are now being made,  at HERA as well as at CERN (COMPASS)
 and Brookhaven National
Laboratory (STAR and PHENIX at RHIC)  to see if two asymmetries are related
and also to measure the polarization of
the glue. 

Clearly one also needs a matching theoretical efforts to get 
a dynamical explanation to all these phenomena, as the pQCD cascade
(which is flavor and (approximately) chirality-blind) obviously cannot
provide.

   One particular direction of such studies are related with
the non-perturbative phenomena in the QCD vacuum described
semiclassically
by $instantons$. There are several qualitative arguments why instantons
may be clue to a potential explanation of these puzzles.
Forte and Shuryak \cite{Forte:1990xb}
had argued that instantons provide a 100 per cent
effective mechanism of a polarization transfer from quarks to gluons. 
Kochelev \cite{Kochelev:1995vy} noticed that
a sea produced via 't Hooft vertex from say left-handed $u_L$ valence quark is 100
per cent flavor polarized, as it can only have $\bar d_L d_R,\bar s_L s_R$
pairs,
and also should have the opposite chirality.
Unfortunately, none of these ideas so far resulted in some quantitative
predictions, see e.g. a recent work by T.Shchafer and Zetocha
\cite{Schafer:2004ke} on the spin crisis.

One more, although indirect, argument came from lattice calculations.
 Negele et al (the MIT group)
  \cite{Negele:2004iu}  have noticed  that
moments of the various structure functions change very little when
the true ``quantum'' lattice
configurations are substituted by ``semiclassical''
(or ``cooled'') ones. This procedure, which eliminates pQCD
gluons and most of quantum fluctuations from vacuum configurations,
is known to
preserve mostly the instantons,
reasulting basically in configurations of the ``instanton liquid''.
If true,  this observation suggests that instantons alone would be
sufficient to derived all structure
functions, including the spin-dependent ones.

 The
Single Spin Asymmetries
(SSA) are large spin-dependent effects which are under intense experimental
study. 
{So far their theoretical discussion  (see e.g.\cite{Mulders:1995})
have aimed mostly at their proper
parameterization rather than explanation. One importnant step was an
 introduction of the nontrivial T-odd structure in the initial
state via appropriate structure function is called the Sivers effect
\cite{Sivers:1990fh,Sivers:1989cc}, while a similar effect
in fragmentation function is called
the Collins effect \cite{Collins:1992}.
 The corresponding function was introduced in \cite{Boer:1997nt}. 
In both cases the hard block remains the usual lowest-order pQCD scattering.
One more logical alternative is the {\em next twist} hard collision,
in which at the moment of hard scattering there is extra gluonic field
(or a gluon), which can be also correlated with the spin \cite{Qiu:1998ia}.

  At a deeper level,  two issues have been singled out, to be included 
as the key ingredients of any
theoretical explanation of the asymmetry, as they are
  both the
necessary prerequisites to the very existence of SSA. Those are (i) the
  {\em chirality flip}; and (ii) the {\em final state interaction} of the
outgoing quark.
These issues are best explained if the state of the transversely polarized
nucleon is viewed as a superposition of plus and minus chirality states.
SSA can only result from the interference of the two amplitudes, while
the usual pQCD handbag diagram conserves chirality of the quark.

The first issue can be satisfied e.g. by the introduction of a new
component of the nucleon
 wave function, in which the valence quark rotates orbitally
  and thus has a spin opposite to that of the nucleon. 
The second is related to the
 decade long theoretical stalemate over Sivers effect. Namely, 
Collins \cite{Collins:1992} have
argued that it should be zero based on T-invariance. The
proof was retracted later, and the loophole is precisely the P-exponents
of the outgoing quarks, or their possible final state interaction. 

 Brodsky, Hwang, and 
Schmidt (BHS)
\cite{Brodsky:qv,Brodsky:2002cx}, and Collins \cite{Collins:2002} 
have then shown how the Sivers effect could be incorporated into the 
QCD framework. For early model of a T-odd {\em distribution} 
function see\cite{Bacch1}, 
as well as a bag-model calculation by Yuan\cite{Yuan:2003wk},
and a  model with spin-0 AND spin-1 
diquarks in\cite{Bacchetta:2003rz}.

BHS used a very simplified model, in which a nucleon is made of a valence
quark, which carry all the angular momentum, plus the spin-zero diquark.
The issue (i) is included via new p-wave wave function, and (ii) via
the lowest order gluon exchange between the outgoing quark and
the rest of the system (the diquark). Note that in BHS approach
there is no connection
between (i) and (ii): just the final state interaction is necessary to make
the nontrivial sector of the nucleon wave function visible.

%

The philosophy of
our approach came out of reflections about this very point. We thought
it is quite likely that the underlying dynamics
of the quark chirality flip 
is related with the nonperturbative interaction producing
chiral symmetry breaking. (By the way, throughout this paper we will
ignore nonzero quark masses and thus treat chiral symmetry as exact.) 
in the QCD vacuum.  There are convincing arguments that this phenomenon is
generated by  small-size instantons, see \cite{SS_98} for a review.
And as instantons can provide the chirality flip, they also are capable
to generate
large ($O(1)$ rather than $O(\alpha_s)$)  phase of the P-exponent,
the final state interaction of the struck quark. As both are necessary
for
the asymmetry in question, it makes instanton mechanism twice more
attractive
candidate for its explanation.

In short, the physics of our proposal is as follows:
the asymmetry appears when
the quark-lepton collision point happens to be  close
to a preexisting topological vacuum fluctuation, due to tunneling through the
topological barrier of QCD. 
In this sense the phenomenon is generic, not related to the nucleon
itself,
but rather proportional to density of instantons (tunneling events)
in the QCD vacuum.

 As the initial quark
is moving in a strong color field and can ``disappear into a Dirac
sea'' while
instead another quark, with the {\em opposite} chirality appears close by.
This is accounted by the so called 't Hooft zero mode term in a quark 
propagator.
 Also the final state interaction effect is directly given by a quark 
propagator on top of the instanton background: in fact it
is much enhanced compared to BHS paper because
the gluon field originating from instanton is large $A\sim 1/g$ 
compared to perturbative field of the exhcanged gluon. As a result
there is no extra small factor $\sim\alpha_s$ in the amplitude.

We will thus obtain the resulting asymmetry from well-known propagators
in a relatively simple form, using then  the
vacuum properties from the instanton-liquid model \cite{SS_98} for quantitative
estimates.
Let us emphasise that we does not parameterize the effect via
 introduction of some  unknown parameters or structure/fragmentation
functions, but express
it in terms of relatively well known 
parameters of the vacuum, tested in particular in other applications
of instanton dynamics and/or muliple lattice works.

 HERMES 
experiment have reported first data 
on SSA in SIDIS for longitudinally \cite{Airapetian:2002mf,Airapetian:2001iy,Airapetian:1999tv} 
and transversly polarized targets \cite{Airapetian:2004tw}. 
There is also data from COMPASS \cite{Avakian:2003pk} for
longitudinally polarized targets.
Contributions
of various dependence on the spin direction/transverse momentum
of produced particle directions is possible to disentangle.}


The plan of the paper is as follows: in Section 2 kinematics of SSA in SIDIS
is overviewed, in Section 3 the asymmetric part of the cross section due to
instantons is calculated, in Section 4 an estimate
of the effect and a comparison with experiment are given.     

\section{Kinematics of azimuthal SSA in SIDIS}

Total cross section for deep inelastic scattering  has the form
\be
\frac{d\sigma}{dx dy d\phi}= 
\frac{\alpha_{em}^2}{Q^4}y L^{\mu\nu}W_{\mu\nu} ,
\ee
where azimuthal angle $\phi$ is unobservable in totally
inclusive DIS.

\begin{figure}
 \includegraphics[width=8cm]{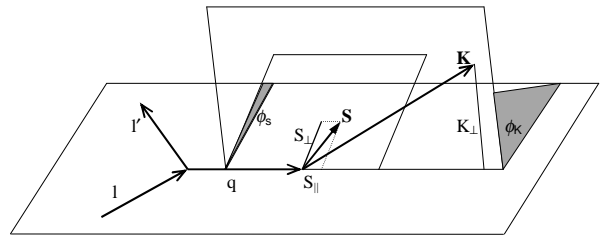}
 \caption{\label{fig_kinem}
Kinematics of single particle inclusive DIS in nucleon
rest frame, defining all momenta and angles to be used.}
 \end{figure}

Symmetric (spin independent) lepton
tensor is given by (see Fig.\ref{fig_kinem})
\be
L^{\mu\nu}=2(l^\mu l^{\prime\nu}+l^{\prime\mu}l^\nu)
-2g^{\mu\nu}(l\cdot l')
\ee

For totally inclusive cross section symmetric
part of $W_{\mu\nu}$ is given by
\ba
W_{\mu\nu}&=&\left(-g_{\mu\nu}+\frac{q_\mu q_\nu}{q^2}
\right)F_1(x,Q^2)  \nonumber \\
&+& 
\left(P_\mu+\frac{1}{2x}q_\mu\right)
\left(P_\nu+\frac{1}{2x}q_\nu\right)
\frac{F_2(x,Q^2)}{P\cdot q}
\ea

In SIDIS one has one more vector parameter, the momentum of the
produced hadron, $K_\mu$. This leads to the appearance of
several new possible tensor structures of hadronic tensor
$W_{\mu\nu}$ and new dimensionless invariants on which 
"structure functions" may depend. The tensor structure
of $W_{\mu\nu}$ is of course limited by symmetry,
$W_{\mu\nu}=W_{\nu\mu}$ (we consider only unpolarized
electrons for leptonic tensor), electromagnetic gauge invariance,
$q^\mu W_{\mu\nu}=0$, and parity invariance.
We are interested also in {\it spin-dependent} asymmetries
and therefore, nucleon spin $S_\mu$ ($S^2=-1$) must be involved in
nontrivial combination with produced hadron momentum.
To limit the possible structures even more, we will
consider hadronic tensor only to the first power in
$K$, assuming that it enters $W_{\mu\nu}$ in the combination
$K/Q$, which is generally small.
 
Then, the possible tensor structures are 
$(P+\frac{1}{2x}q)_{\{\mu}\epsilon_{\nu\}\rho\sigma\tau}q^\rho K^\sigma 
S^\tau$ and $\epsilon_{\pi\rho\sigma\tau}P^\pi q^\rho K^\sigma 
S^\tau$, the latter being multiplied either on 
$(-g_{\mu\nu}+\frac{q_\mu q_\nu}{q^2})$ or on 
$\left(P_\mu+\frac{1}{2x}q_\mu\right)
\left(P_\nu+\frac{1}{2x}q_\nu\right)$

Although $\epsilon^{\pi\rho\sigma\tau}
P_\pi q_\rho K_\sigma S_\tau$ may not be excluded
on general grounds, this structure leads to small
contribution in the parton model. Indeed, in the 
parton model, the dependence of $W_{\mu\nu}$
on $P_\mu$ is possible only through
the momentum of struck quark. In the infinite 
momentum frame $p_\mu=xP_\mu$. 
It is also true that $K_\mu=z k_\mu$, where $k_\mu$ 
is the momentum of the quark after the collision,
up to relatively small correction in the fragmentation process.  
Momentum conservation in the interaction vertex gives
$k_\mu=p_\mu+q_\mu$ up to small correction due to
possibility of rescattering. Overall, it leads to
the extra power of small transverse momentum for 
$\epsilon^{\pi\rho\sigma\tau}
P_\pi q_\rho K_\sigma S_\tau$.

As we will see in the next section, instanton-induced
contribution has the form
\be
W_{\mu\nu}\sim (p+k)_{\{\mu}\epsilon_{\nu\}\rho\sigma\tau}q^\rho k^\sigma 
s^\tau
\ee
in accordance with the general analysis ($s$ being spin of the
quark).

Last but not least: these combinations of 2 momenta, one energy
and one spin is T-odd. Therefore one can only find their
contribution to any observable multiplied by another T-odd quantity,
such as final state interaction phase. 

\section{Quark scattering in instanton field}

Let us consider incoming quark with the momentum
$p$ and density matrix $\not{p}(1+\gamma_5\not{s})$.
Because we are interested only in spin-dependent
part of the cross section we take only $\not{p}\gamma_5\not{s}$ as
the quark density matrix.

For spin-dependent hadron tensor one has
\be\label{Wdef}
\Delta W_{\mu\nu}=\mbox{Re} \left[\mbox{tr}(\hat{k} M^1_{\{\mu}  
\hat{p}\gamma_5\hat{s} M^0_{\nu\}})\right],
\ee
where the zeroth order vertex 
$M^0_\nu=\gamma_\nu$, while the first order
one $M^1_\mu$ is a sum of amplitudes in
instanton and antiinstanton fields, $\hat{k}$ is the density
matrix of out-going quark (matrix elements here defined without 
projections on in- and out- going states).
Two indices with curly brackets are assumed to be symmetrized. 

Spin-dependent part of in-coming quark density matrix is chirally-odd,
therefore, taking to account that all other parts of Eq.~(\ref{Wdef})
are chirally-even, $M^1_\mu$ must be chirally-odd. Therefore, $M^1_\mu$
must contain propagation through zero-mode in instanton
(antiinstanton) field. Calculation of $M^1_\mu$ is most easily performed
in chiral basis first. We return to the Dirac fermions in the end. 

The calculation of $M^1_\mu$ is in the complete analogy to
the calculation of the instanton-induced
chirally-odd contribution to the gluon structure function
made by Moch, Ringwald, and Schrempp \cite{Moch:1997}, where reader
is referred for all technical details.

In instanton field one has for quark propagator due to zero
mode (x-space, non-amputated, left-to-right flip)
\be\label{x_zero}
S_0(x,y)_{\dot{\beta}}\strut^{\dot{\alpha}}\strut^j_i=
\frac{\rho^2}{\pi^2\lambda}
\frac{x_\gamma \bar{\sigma}^\gamma_{{\dot{\beta}} \rho} \epsilon^{\rho j}
x_\delta (\sigma^\delta)^{{\dot{\alpha}} \pi} \epsilon_{i \pi}}{(x^2+\rho^2)^{3/2}(y^2+\rho^2)^{3/2}|x||y|}
\ee
Here Greek indicies are Weyl spinor indicies, $i$ and $j$ are color
indicies, $\sigma_\mu=(i, \vec{\sigma})$, $\bar{\sigma}_\mu=(-i, \vec{\sigma})$
(Euclidean space), $\epsilon^{01}=-\epsilon^{10}=-\epsilon_{01}=\epsilon_{10}=1$
are projectors on zero mode chiral-color states. Zero mode propagator
is normalized to $\lambda$, the lowest eigenvalue of the Dirac operator
in instanton liquid. We return to the discussion of its value in Section 4.

Fourier transform with respect to incoming particle is given by
\be\label{p_zero}
S_0(x,p)_{{\dot{\beta}} \alpha}\strut^j_i=
\frac{2\rho^2}{\lambda} \frac{1}{(x^2+\rho^2)^{3/2}}
\frac{x_\gamma \bar{\sigma}^\gamma_{{\dot{\beta}} \rho} \epsilon^{\rho j}
\epsilon_{i \alpha}}{|x|},
\ee
where mass-shell condition ($p^2=0$) is assumed and incoming particle
propagator is amputated.

Non-zero mode propagator for {\it right-handed} quark, 
this is right-handed after
the flip on zero mode, is (Fourier transformed and amputated for 
outgoing particle)
\ba
S_{nz}(k,x)^\beta\strut_\alpha\strut^j_i&=&-\frac{|x|}{\sqrt{x^2+\rho^2}}e^{ik\cdot x}
\delta^\beta_\alpha [\delta^j_i \nonumber\\ 
&+& 
\frac{\rho^2}{x^2}\frac{(\bar{\tau}_\rho \tau_\sigma)^j_i k^\rho x^\sigma}{2 k\cdot x} 
\left(1-e^{-ik\cdot x}\right)
], \label{p_nonzero}
\ea
where $\tau_\mu=(i, \vec{\tau})$, $\bar{\tau}_\mu=(-i, \vec{\tau})$ 
for color matricies.

We now can see the T-odd phase, which enters the propagator of the
quark in instanton field. For the purpose of demonstration we will
drop $\exp(-ik\cdot x)$ from round brackets in Eq.~(\ref{p_nonzero}).
It serves to give to the propagator correct $x\rightarrow 0$ limit.
One can think that we are working in $k\cdot x \gg 1$ kinematical
domain. Then, Eq.~(\ref{p_nonzero}) becomes
\be
S_{nz}(k,x)\simeq -e^{ik\cdot x}
\exp\left[i\frac{\bar{\eta}^{a\mu\nu}k_\mu x_\nu \tau_a}{k\cdot x}
\ln\left(\frac{x^2+\rho^2}{x^2}\right)\right],
\label{phase}
\ee
where $\bar{\eta}^{a\mu\nu}$ is 't Hooft symbol. It makes the phase
T-odd after continuation to Minkowski space.
 
Contribution to $M^1_\mu$ is given by
\be
M^1_\mu=\int d^4x e^{iq\cdot x} S_{nz}(k,x)\sigma_\mu S_0(x,p)
\ee
Before calculating the integral one has to perform some algebra. Namely,
chiral-color projectors of zero modes effectively mix chiral ($\sigma$) and
color ($\tau$) matricies. We make use of the relation
\be
\epsilon_{\gamma \alpha} (\sigma_\mu)^{\gamma\dot{\gamma}} 
(\bar{\sigma}_\nu)_{\dot{\gamma} m} = 
\epsilon_{\gamma m} (\sigma_\nu)^{\gamma\dot{\gamma}} 
(\bar{\sigma}_\mu)_{\dot{\gamma} \alpha}
\ee
Therefore, there is no much sense of keeping difference between 
$\sigma$ and $\tau$ matricies, when required we will indicate 
correct indicies to distinguish between color and 
chiral degrees of freedom.

After integration
\be
M^1_\mu=\frac{4\pi\rho^2}{\lambda} i (\sigma_\mu\bar{\sigma_\rho}\epsilon)^{\alpha i}
\epsilon_{\beta j} \Phi^\rho(k,q)
\label{flip-first}
\ee
with
\be
\Phi^\rho(k,q)=-\frac{p^\rho}{p^2}-\frac{k^\rho}{2k\cdot q}(1-f(\rho |q|)),
\ee
where $f(a)=aK_1(a)$ ($f(0)=1$), and we took into account that $k=p+q$. First term in the expression
of $\Phi^\rho$ contains 0 in the denominator, however in the nominator one effectively
has $\bar{\sigma}_\rho p^\rho$, which being multiplied on in-coming quark state
gives 0. This term does not contribute to the spin-dependent part of the cross section anyway,
so we may neglect it altogether.
\vskip .5cm
 \begin{figure}
 \includegraphics[width=8cm]{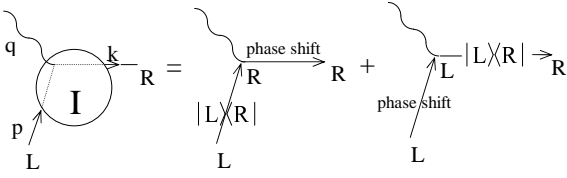}%
 \caption{\label{fig_diagram}
The amplitude for a single quark scattering in an instanton background
field (shown schematically by a circle with I)
can be written as a sum of two diagrams. They differ by where the
chirality flip,
between a chosen left-handed (L) initial struck quark into a
right-handed (R) one, which is described by the zero mode part
of the propagator. The ``phaseshift'' subscript reminds about 
a complex phase of the non-flip part of the propagator, which has to
be kept to get the nonzero answer.
}
 \end{figure}

Eq.~(\ref{flip-first}) corresponds to applying zero mode propagator to
the in-coming quark before its collision with a virtual photon. There is,
of course, another diagram (see Fig.\ref{fig_diagram}), which refers to zero mode propagator inserted
in the out-going quark line. That can easily be found by Hermitian conjugation
of Eq.~(\ref{flip-first}) and substitution $k\leftrightarrow -p$, which gives
\be
M^{1\prime}_\mu=-\frac{4\pi\rho^2}{\lambda} i \epsilon_{i \alpha}
(\epsilon\sigma_\rho\bar{\sigma_\mu})^{j \beta}
\Phi^\rho(p,q)
\label{flip-last}
\ee

Next, we are taking trace over color indicies. Because all other parts of
the diagram are trivial in color, one reduces $M^1_\mu$ to
\be
M^1_\mu=\frac{4\pi\rho^2}{\lambda} i (\sigma_\mu\bar{\sigma_\rho} \Phi^\rho(k,q)
-\sigma_\rho\bar{\sigma_\mu} \Phi^\rho(p,q))
\label{color-traced}
\ee

The matrix element for propagation in the anti-instanton field is given
by $\bar{\sigma}\leftrightarrow\sigma$.

Inserting $M^1_\mu$ to the Eq.~(\ref{Wdef}) one has
\ba
\Delta W_{\mu\nu}&=&\frac{4\pi\rho^2}{\lambda Q^2} \mbox{Im} [
\mbox{tr}(\hat{k} \hat{p} \gamma_{\{\mu} \hat{p}\gamma_5\hat{s}
 \gamma_{\nu\}}) \nonumber \\ 
&+&
\mbox{tr}(\hat{k} \gamma_{\{\mu} \hat{k} \hat{p}\gamma_5\hat{s} \gamma_{\nu\}})
] (1-f(\rho |q|))
\label{Wall}
\ea

Some trivial Dirac algebra gives
$$
\Delta W_{\mu\nu}=\frac{8\pi\rho^2}{\lambda} \frac{1}{Q^2} (k + p)_{\{\mu}\mbox{Im} 
\,\mbox{tr}(\hat{k} \hat{p} \gamma_5\hat{s} \gamma_{\nu\}})
(1-f(\rho |q|))
$$
\be
=-\frac{32\pi\rho^2}{\lambda} \frac{1}{Q^2} 
(k + p)_{\{\mu}\varepsilon_{\nu\}\rho\sigma\tau}q^\rho k^\sigma s^\tau (1-f(\rho |q|)).\label{parton_answer}
\ee
In the last expression we switched back to Minkowski space and made use
of $\sigma_0^E=i \sigma_0^M$.

In this calculation we have neglected  the interaction between instanton
and the rest of the nucleon, apart of the struck
quark. This aproximation is
motivated
by the fact that the typical instanton size in the QCD vacuum
$\rho\approx 1/3 fm$ is small compared to nucleon size $R_N$.
Their account would lead to corrections of the order $\rho^2/R_N^2\sim 1/10$. 

In vacuum parametrized by an ensemble of instantons one has to 
integrate Eq.~(\ref{parton_answer})
over collective degrees of freedom: color rotations and instanton
size. 
\be
\overline{\left(\frac{\rho^2}{\lambda}\right)}
\rightarrow \frac{\kappa}{\bar{\rho}^2m^*},
\ee 
where $\kappa$ is the instanton diluteness factor, $\bar{\rho}^2$
is the characteristic instanton size, and $1/m^*$ is the inverse
effective quark mass in the instanton-liquid model.

\section{Estimate of the asymmetry}

Eq.~(\ref{parton_answer}) constitutes the result for spin-dependent
asymmetric tensor at partonic level. To change it to hadronic
result one has to substitute $p=xP$, $k+p=2xP+q$, and $k=K/z$.

To find out the correct kinematical normalization we compare analogous
calculation for spin-independent DIS cross section, which gives
\be
W^{parton}_{\mu\nu}=
2N_c(p_\mu k_\nu + p_\nu k_\mu - g_{\mu\nu} (p\cdot k)).
\ee 
Rewriting this through conventional structure functions one finds
that correct normalization is given by multiplication of 
$W^{parton}_{\mu\nu}$
on 
\be
\sum\limits_q e^2_q f_q(x)/(2 N_c Q^2)
\ee 
and we have $F_1(x)=F_2(x)/(2x)=\frac 12 \sum\limits_q e_q^2 f_q(x)$ 
for conventional structure functions in our approximation.
Because of the spin dependence of Eq.~(\ref{parton_answer}) one
has to use spin-dependent quark 
distributions, which we tentatively will call $f_{q,s}(x)$.
We can now rewrite Eq.~(\ref{parton_answer}) as
\ba
\Delta W_{\mu\nu}&=&\frac{32\pi\kappa}{\bar{\rho}^2m^*N_c} \frac{x}{z} 
\frac{1}{Q^4} (P+(1/2x)q)_{\{\mu}
\varepsilon_{\nu\}\rho\sigma\tau}q^\rho K^\sigma s^\tau \nonumber\\
&\times&
(1-f(\rho Q))\sum\limits_q e^2_q f_{q,s}(x) D_q(z).\label{hadron_answer1}
\ea

Asymmetric part of the cross-section is now 
\ba
\frac{d\Delta\sigma}{dx dy dz d\phi_K}= \frac{\alpha_{em}^2}{Q^2}
\frac{32\pi\kappa}{\bar{\rho}^2m^*N_c}
\frac{|K_\perp|}{z Q}(1-f(\rho Q))&&\nonumber\\ 
\times\sum\limits_q e^2_q f_{q,s}(x)D_q(z) 
\left(\frac{2}{Q}\frac{1-y}{y}\sin(\phi_K-\phi_s)|s_\perp|\right.&&\nonumber\\
+\left.\frac{(1-y/2)\sqrt{1-y}}{Mx}\sin\phi_K s_\parallel\right).&&
\label{delta_sigma1}
\ea
One can now define what exactly are the spin-dependent quark
distributions $f_{q,s}(x)$ we introduced in analogy to $f_q(x)$.
They correspond to the probability to find a quark in a hadron
polarized the same way as a hadron minus the probability
to find a quark polarized in opposite direction then a hadron.
Schematically,
\be
f_{q,s}(x)s \rightarrow \Delta f(x) s_\parallel + \Delta_T f(x)
s_\perp.
\ee
Inserting it in Eq.~(\ref{delta_sigma1}) gives
\ba
\frac{d\Delta\sigma}{dx dy dz d\phi_K}= \frac{\alpha_{em}^2}{Q^2}
\frac{32\pi\kappa}{\bar{\rho}^2m^*N_c} \frac{|K_\perp|}{z Q}
(1-f(\rho Q))&&\nonumber\\ 
\times\sum\limits_q e^2_q D_q(z)
\left(\frac{2}{Q}\frac{1-y}{y}\sin(\phi_K-\phi_S)|S_\perp|\Delta_T f_q(x)
\right.&&\nonumber\\
\left.
\frac{(1-y/2)\sqrt{1-y}}{Mx}\sin\phi_K S_\parallel \Delta f_q(x)
\right).&&\label{delta_sigma2}
\ea

To obtain relative asymmetry one has to compare Eq.~(\ref{delta_sigma2})
to totally inclusive cross section
\be
\frac{d\sigma}{dx dy dz d\phi}= \frac{\alpha_{em}^2}{Q^2}\frac{1+(1-y)^2}{y}
\sum\limits_q e^2_q f_{q}(x)D_q(z)\label{sigma_tot}
\ee

From Eqs.~(\ref{delta_sigma2}) and (\ref{sigma_tot})
one can see that for the most simplistic model of a nucleon,
where total spin and charge are carried out by single quark
(the other two being insulated from virtual photon in diquark state),
relative asymmetry does not depend on distribution functions
of the nucleon and is in a sense universal, applicable to {\em 
all other hadrons}. 

More realistic approximation is $\Delta f(x)=\Delta_T f(x)$.
It ignores differences due to relativistic motion of the
quarks inside nucleons. However, in absence of reliable
experimental data on $\Delta_T f(x)$ one can use this approximation
to get reasonable estimate of the transverse asymmetry.
Model calculations also favor such an approximation. 

From Eq.~(\ref{delta_sigma2}) one can readily see that if
$\Delta f(x)=\Delta_T f(x)$ is assumed, the relative size
of transverse and longitudinal asymmetries is purely
kinematical and does not depend on any details of hadronic
structure. 

We will now give an estimate for prefactor in Eq.~(\ref{delta_sigma2})
from the single instanton approximation 
(SIA) of instanton-liquid model. 
For general discussion of instanton phenomenology the reader can
consult e.g. \cite{SS_98}. We will use the usual diluteness parameter
and size
\be  \label{eq_liquid}
\bar\rho=1/3\, fm \hspace{1cm} \kappa=n\bar\rho^4\approx 1/3^4 \ee
As for the accuracy of SIA and
the value of the (apropriately averaged) 
value of the Dirac eigenvalues $m^*$, see detailed
discussion in ref.\cite{SIA}. It is found there that if it would be
simply a quantity with one zero mode, like $<\bar q q>$, the 
accuracy of selecting one closest instanton from the ensemble
and ignoring all others is typically about 30\%. In this case
the definition of it (called $m_{uu}$ in \cite{SIA})  should be
 $m^*\equiv(<1/\lambda>)^{-1}$ where the angular bracket stands for
real eigenvalue spectrum in the vacuum ensemble. 
Its numerical value changes from $m^*=120 \, \MeV$ for random instanton
liquid model to $m^*=170 \, \MeV$ in interacting instanton ensemble.
It must be noted that in our
calculation spin asymmetry depends on both chirality flip
and phase shift on the same instanton. Thus, we expect that 
in this case SIA is more accurate and use $m^*=170 \, \MeV$.
In summary, all instanton-related parameters appear in the following
combination, which has the dimention of the energy
\be
\frac{32\pi\kappa}{\bar{\rho}^2m^*N_c}=0.88\, \GeV
\ee
Although it makes a parameter of the order of 1 GeV, one should keep
in mind that it includes the instanton density which  is
nonperturbatively
small $\kappa\sim exp(-2\pi/\alpha_s(\rho))$. Furthermore, the
phenomenological smallness (\ref{eq_liquid}) is not seen
only because it happens to be compensated by large
numerical
factor $32\pi$. 

\subsection{Comparison with experiment}

Detailed comparison with the experiment is outside
the scope of this paper. We present here only a few details
to establish phenomenological relevance of our model.
We consider longitudinal and transversal spin asymmetries for production of $\pi^+$ mesons
off polarized proton target \cite{Airapetian:2002mf,Airapetian:2001iy,Airapetian:1999tv}. 
For simplicity, we will assume 
that in order to produce $\pi^+$ from the proton one has
to struck a $u$ quark. In other words, $D^{\pi^+}_q(z)=0$ unless
$q=u$. Then, from Eqs.~(\ref{delta_sigma2}), (\ref{sigma_tot})
longitudinal asymmetry is
\ba
A^{\sin\phi}_{UL}&=&0.88\, \GeV \frac{|K_\perp|}{z Q}
(1-f(\rho Q))
\frac{y(1-y/2)\sqrt{1-y}}{M (1+(1-y)^2)} \nonumber\\
 &\times&
\frac{\Delta f_u(x)}{x f_u(x)}.\label{AUL}
\ea

 \begin{figure}
 \includegraphics[width=8cm]{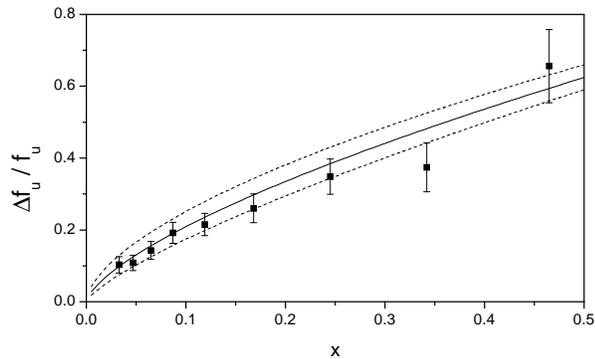}%
 \caption{\label{fig_polar}
 Relative polarization of $u$-quark in the proton.
The error bars combine statistical and systematic errors. Parametrization
$x^\alpha$ with $\alpha=0.68\pm 0.08$ is shown by solid (dashed) lines.} 
\end{figure}

\begin{figure}
 \includegraphics[width=8cm]{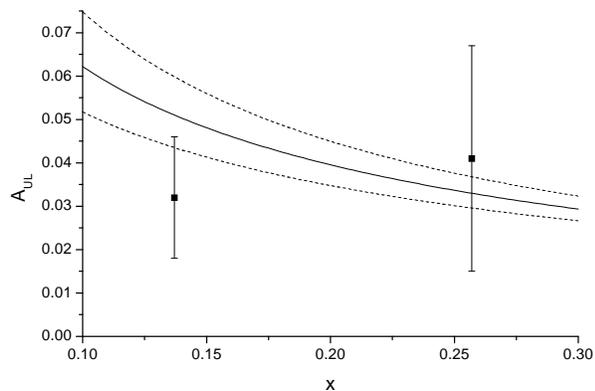}%
 \caption{\label{fig_asym}
 Experimental values $A_{UL}$ for moderate 
$x$ are shown with comparison with the model prediction. Theoretical 
uncertainty is due to uncertainty in polarized distribution function.}
 \end{figure}

The ratio of polarized to unpolarized distribution function
is measured by HERMES collaboration 
\cite{Ackerstaff:1999,Airapetian:1998,Ackerstaff:1997} for the same
kinematical region as spin asymmetries. It is shown on
Fig.\ref{fig_polar}. It may be fitted with reasonable accuracy by simple
power law $\Delta u/u=x^{\alpha}$ with $\alpha=0.68\pm 0.08$.
(Note that this dependence should not be true down to very low $x$, or
else the $\Delta u/(x u)$
blows up.) 
Parameters $Q^2$, $x$, and $y$ are related by $Q^2=xy(s-M^2)$
(here $s$ is Mandelstam variable, $s=2ME$ in proton rest frame).
$|K_\perp|$ and $z$ can be taken as independent from the rest
of kinematical variables as long as $|K_\perp|/z\ll Q$. Otherwise
DIS separation of parallel and transversal degrees of freedom
breaks down. In HERMES experiment $\langle K_\perp \rangle=0.44$ and
$\langle z\rangle=0.48$, while $Q^2$ is constrained to be $>1\GeV^2$.
Thus, we assume $|K_\perp|/z=0.92\GeV$ in Eq.~(\ref{AUL}).
Then we average over $0.2<y<0.85$. The result for moderate values
of $x$ is shown in Fig.\ref{fig_asym}. We have excluded $x<0.1$ because
for small $x$ our simplifying assumptions about proton structure
are not applicable.

\begin{figure}
 \includegraphics[width=8cm]{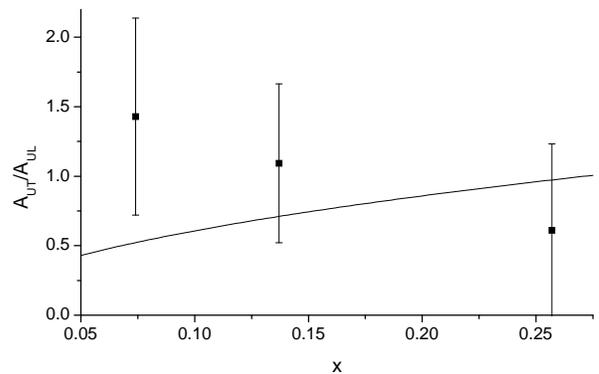}%
 \caption{\label{fig:relative}
 Experimental values of relative transversal to longitudinal
asymmetries $A_{UT}/A_{UL}$ for moderate 
$x$ are shown with comparison with the model prediction. }
 \end{figure}

Relation of transversal to longitudinal asymmetries for the same
simplified model of $\pi^+$ production we use is
(see Eq.~(\ref{delta_sigma2}))
\be\label{relative}
\frac{A_{UT}}{A_{UL}}=\frac{2\sqrt{1-y}\sqrt{x}}{(1-y/2)y^{3/2}}  
\frac{M}{\sqrt{s-M^2}}\frac{\Delta_T f_u(x)}{\Delta f_u(x)}
\ee
Recall that in our simplified model $\Delta_T f_u(x)=\Delta f_u(x)$.
Taking to account HERMES kinematics as outlined above one
finally has an estimate
\be
\frac{A_{UT}}{A_{UL}}=1.92\sqrt{x}
\ee
which is compared to the available data  in Fig.~\ref{fig:relative}

\section{Conclusions and Outlook}

In this paper we have made a step toward the semicalssical
 theory of
various spin-dependent effects in QCD, based on instantons.

The advantages of the instanton-induced mechanism as an
explanation of azimuthal spin asymmetris are based on 
the fact that they provide simultaneously both requied ingredients.
First,  instantons are the well known  source of the chirality flip,
relevant in the kinematic domain of the
scale of about 1 GeV. Second,  instantons also provide large T-odd
phase for the outgoing quark.
We emphasize that the use of instantons allows us not to
introduce any new parameter or structure/fragmentation functions,
but express the result in terms of well known quantities.
The main of them is the ``vacuum diluteness'' parameter 
(\ref{eq_liquid}), which gets compensated by a large numerical factor $32\pi$
in the answer.

The magnitude of the effect is thus fixed with the 
 absolute normalization (\ref{AUL}), based on the
 parameters of the instanton ensemble model known since 1982, see \cite{SS_98}.
The result agrees in sign and magnitude with the available experimental data
in suitable kinematic domain.
We have argued that the asymmetry does not depend on the specific
 distribution functions of the nucleon, and is thus universal to all 
other hadrons.

Furthermore, our
 spin-dependent azymuthal asymmetries
have a particular tensor structure in the lowest nonzero order of
$K_\perp/Q$ as long as parton interpretation of hadron structure
is taken into account. It leads to the specific prediction for the 
dependence of longitudinal and transverse asymmetries 
on kinematical parameters which is completely
independent on the phenomenological imputs.

 For the outlook, one may think that  the explanation of other spin
asymmetries, e.g. in {$pp\uparrow$} collisions, can also be provided
by instantons. 
  The FERMILAB data 
\cite{Adams:1991cs,Adams:1991rw,Bravar:1996ki}
revealed considerable asymmery in pion production starting from $x_F\sim 0.5$ 
and rising towards higher values of $x_F$. 
The explanation of this asymmetry based on instanton 
mechanism was pioneered by Kochelev \cite{Kochelev:1999nd,Kochelev:1997fv} who provided qualitative expressions for it.
More quantitative calculation would however be needed to relate this
effect
to spin effects in DIS we dicuss above.

One more direction of future  work
 may  be combined with the description of
even the $non-polarized$ DIS  in the $x_F\rightarrow 1$
limit, where it is known to be dominated by large
 higher twist effects. Those were also speculated long ago
 to be due to instantons
\cite{SV_twist}, but it was never demonstrated.
 If that conjecture happens to be true, the instanton diluteness $\kappa$
 would drop out from numerator and denominator of 
 the spin asymmetry, resulting in really large $\sim O(1)$ and truly 
universal  
asymmetry independent on $\kappa$.

\section*{Acknowlegements}
One of us (ES) thank S.Brodsky for useful conversation on the subject.
We are also grateful to A.Bacchetta for his helpful comments 
on the first version of the manuscript.
The work is partialy supported by US-DOE grants 
DE-FG02-88ER40388 and DE-FG03-97ER4014.

\end{document}